\documentclass[journal]{IEEEtran}

\usepackage{amsmath}
\usepackage{subfigure}
\usepackage{graphicx}
\usepackage{latexsym}

\usepackage{multirow}
\usepackage{amsmath}
\usepackage{stfloats}
\usepackage{xcolor}
\usepackage{amssymb,amsbsy,verbatim,array}
\usepackage{epstopdf}
\usepackage{cite}
\usepackage{color}
\usepackage{empheq}
\usepackage{cases}
\usepackage{color}
\usepackage{bm}
\usepackage{algorithm}
\usepackage{algpseudocode}
\usepackage{xcolor}
\usepackage[colorlinks,
            linkcolor=red,
            anchorcolor=blue,
            citecolor=green
            ]{hyperref}
\begin{document}
%
\title{Battery-constrained Federated Edge Learning in UAV-enabled IoT  for B5G/6G Networks}  
%
%
%
\author{Shunpu Tang, Wenqi Zhou, Lunyuan Chen, Lijia Lai, Junjuan Xia and Liseng Fan
\thanks{S. Tang, W. Zhou,  L. Chen, L. Lai, J. Xia and L. Fan are all with the School of Computer Science, Guangzhou University, Guangzhou, China (e-mail: \{tangshunpu, 2112006156, 2112019037, 2112006122\}@e.gzhu.edu.cn, xiajunjuan@gzhu.edu.cn, lsfan2019@126.com).}
}

\maketitle

\begin{abstract}
 In this paper, we study how to optimize the federated edge learning (FEEL) in UAV-enabled Internet of things  (IoT) for B5G/6G networks, from a deep reinforcement learning (DRL) approach. 
 The federated learning is an effective framework to train a shared model between decentralized edge devices or servers without exchanging raw data, which can help protect data privacy. 
 In UAV-enabled IoT networks, latency and energy consumption are two important metrics limiting the performance of FEEL. Although most of existing works have studied how to reduce the latency and improve the energy efficiency, few works have investigated the impact of limited batteries at the devices on the FEEL. 
 Motivated by this, we study the battery-constrained FEEL, where the UAVs can adjust their operating CPU-frequency to prolong the battery life and avoid withdrawing from federated learning training untimely. 
 We optimize the system by jointly allocating the computational resource and wireless bandwidth in time-varying environments.
 To solve this optimization problem, we employ a deep deterministic policy gradient (DDPG) based strategy, where a linear combination of latency and energy consumption is used to evaluate
 the system cost. Simulation results are finally demonstrated to show that the proposed strategy outperforms the conventional ones. In particular, it enables all the devices to complete all rounds of FEEL with limited batteries and meanwhile reduce the system cost effectively.
\end{abstract}

\begin{IEEEkeywords}
UAV, federated learning, latency, energy consumption, mobile edge computing.
\end{IEEEkeywords}

%
\IEEEpeerreviewmaketitle

\section{Introduction}
\IEEEPARstart{S}{warms} of unmanned aerial vehicles (UAVs) play a non-negligible role in Internet of things (IoT) for beyond the fifth-generation (B5G) and the forthcoming sixth-generation (6G) wireless mobile networks\cite{UAV1,UAV2,UAV3}. 
Due to the mobility and flexibility, UAVs can effectively collect data and  communicate in the edge computing-based IoT networks. Thanks to these advantages, UAVs have been widely used in many application scenarios, such as environmental monitoring, emergence communication, transportation control and remote sensing.

In recent years, deep learning has shown its great success on speech recognition, computer vision and many other domains.\cite{nature_deeplaerning}.
Especially in the era of big data, millions or billions of data are collected by various sensors and are applied to train the deep learning model. The authors in \cite{imagenet} proposed a large-scale hierarchical database for image classification and promoted the emergence of state-of-the-art (SOTA)
CNN models\cite{alexnet,resnet,densenet}. Large amounts of high-quality data are the key to improve the performance of the deep learning model. As a matter of fact, data collection faces a series of challenges. Though a lot of data are produced by heterogeneous devices, especially 
IoT devices and smartphones in the mobile edge, these data are fragmented and scattered in various devices. This brings out a huge communication cost to 
centralize data on the server to train models when we adopt the traditional centralized training method of deep learning. Meanwhile,
increasing concerns of sensitive privacy make data collection more difficult. People do not want their personal data (e.g. figs, voice, chat history) to be uploaded to others’ servers. There are also some companies, banks and hospitals with a lot of sensitive data which are not allowed to divulge. Moreover, laws on privacy protection have been promulgated continuously. Due to these reasons, data are difficult to be centralized to train deep learning models, which is called  ``data island" \cite{yangqiang_fl}. In this context, federated learning has been proposed to break the ``data island" and it makes full use of each device's data to train a model with a fine performance. In the federated learning, all devices use local data to train a model and share the model on the premise of safety to aggregate a global model with the federated optimization\cite{FL2016,average1,FL_opti}.

In the edge computing-based IoT networks, in order to speed up the process of training and inference of deep learning,
a new concept ``Edge AI" was presented to bring model closer to the places where data are generated when the computational capabilities of edge devices
are continually growing\cite{shi_edge,xuchen_Edge_AI_1,xuchen_Edge_AI_2}. Federated learning is also applied to train deep learning models cooperatively 
in MEC networks, which is referred to as federated edge learning (FEEL)\cite{FL_mec,guo2020feel}. Google Inc\cite{google_applied} used an edge server to 
build a keyboard input prediction model based on FEEL, and FEEL can be also applied to improve the performance of content caching without 
gathering users' data centrally for training\cite{content_caching,cui_FL}.

Although the FEEL has been successfully applied to many application scenarios, there still exist some challenges. One major challenge is that there is a high requirement for the latency and energy consumption in UAV-enabled IoT networks, which is one of the important and critical issues. More importantly, UAVs are powered by limited batteries
and they can only work for a few dozen minutes. It is dangerous for UAVs to run out of power when they are working and it is of vital importance to control the remaining power. On the other hand, enough participants are the guarantee of the performance for the FEEL training. With a limited battery power, how to complete more training rounds of federated learning is an unconsidered issue and it is our motivation to control batteries effectively to prolong their service life.
Accordingly, in this paper we study the battery-constrained FEEL, where the UAVs can adjust their operating CPU-frequency to prolong the battery life and avoid withdrawing from federated learning training untimely. We optimize the system by jointly allocating the computational resource and wireless bandwidth in time-varying environments. To solve this optimization problem, we employ a deep deterministic policy gradient (DDPG) based strategy, where a linear combination of latency and energy consumption is used to evaluate
the system cost. The main contributions of this work are summarized below.

\begin{itemize}
  \item We study the resource allocation strategy for the FEEL in complicated scenarios, where UAVs in edge computing-based IoT networks are powered by limited batteries.
        Moreover, the characteristics of each device are different and complicated, such as
        computational capabilities and channel conditions which are time-varying. 
  \item We propose a resource allocation strategy based on the deep reinforcement learning for the FEEL optimization, where the total latency and energy consumption can be minimized by adjusting 
  the CPU-frequency of devices and upload wireless bandwidth for each device. More importantly, the strategy can prolong the battery life of each device and enable
  devices to complete the required rounds of federated learning.
  
  \item  We conduct simulated experiments to evaluate the performance of the proposed resource allocation strategy and compare the performance with some conventional strategies to demonstrate the superiority of the proposed approach.
\end{itemize}

The rest of this paper is described as follows. Existed and relevant works are introduced in Sec. II. Then, Sec. III describes the system model and formulates the optimization problem, and Sec. IV provides the DDPG-based allocation strategy involving the optimization on computational resource and wireless bandwidth in time-varying environments. After that, Sec. V gives some simulation results and discussions, and we finally conclude our work in Sec. VI.
\section{Related Works}
\noindent\textbf{Mobile Edge computing networks:} With the explosive growth of the number of IoT devices in 5G and 6G era,more and more data are mainly generated in the mobile edge, and mobile edge computing (MEC) is used to process a large quantity of data with a lower latency and energy consumption\cite{mec_survey}. Many researchers focus on the offloading strategy in MEC to reduce system cost  (e.g. Zhao\cite{zzc_tii} 
optimized the computation offloading based on the discrete particle swarm algorithm, the authors in \cite{liang_multiuser} studied the multi-user computational offloading 
and Li\cite{lichao} proposed a DRL based approach to solve the problem of offloading for multiuser and  multi-CAP MEC networks and so on \cite{twc_Allocation,tvt_allocation,tcom_allocation})
\newline
\noindent\textbf{FL in the wireless networks:} In MEC networks, there are many existing works about reducing 
the latency and energy consumption by allocating the system resources. But federated learning 
is basically distributed and heterogeneous, and it needs to synchronize all the participants with different channel conditions and 
computational capabilities \cite{towards_FL}. Hence, another challenge is how to reduce the cost of FEEL. In this direction, Takayuki \cite{nishio_client} proposed
a client selection scheme for federated learning in mobile edge to accelerate the training. In \cite{shi2020device,shi2020joint}, the 
authors studied the federated learning in wireless networks and tried to allocate more bandwidth to nodes with poor channel 
or weak computational capabilities.  Due to the lack of continuous connection between the UAV swarms,
a joint powered allocation design was proposed to optimize the convergence of federated learning \cite{uav_fl}. The authors in \cite{ICCC} presented worker-centric model selection for federated learning in the MEC networks. Zhou\cite{blockchain} proposed a blockchain-based FL framework in the B5G network.
So far, few works have considered that most of the IoT devices are powered by limited batteries. Zhan et al. \cite{zhan2020experience} executed DRL to adjust the CPU-frequency
making a good trade-off between the latency and energy consumption. But this work did not quantify the amount of battery power.
  
  \begin{figure}[tbp]
    \centering
    \includegraphics[width=3.5in]{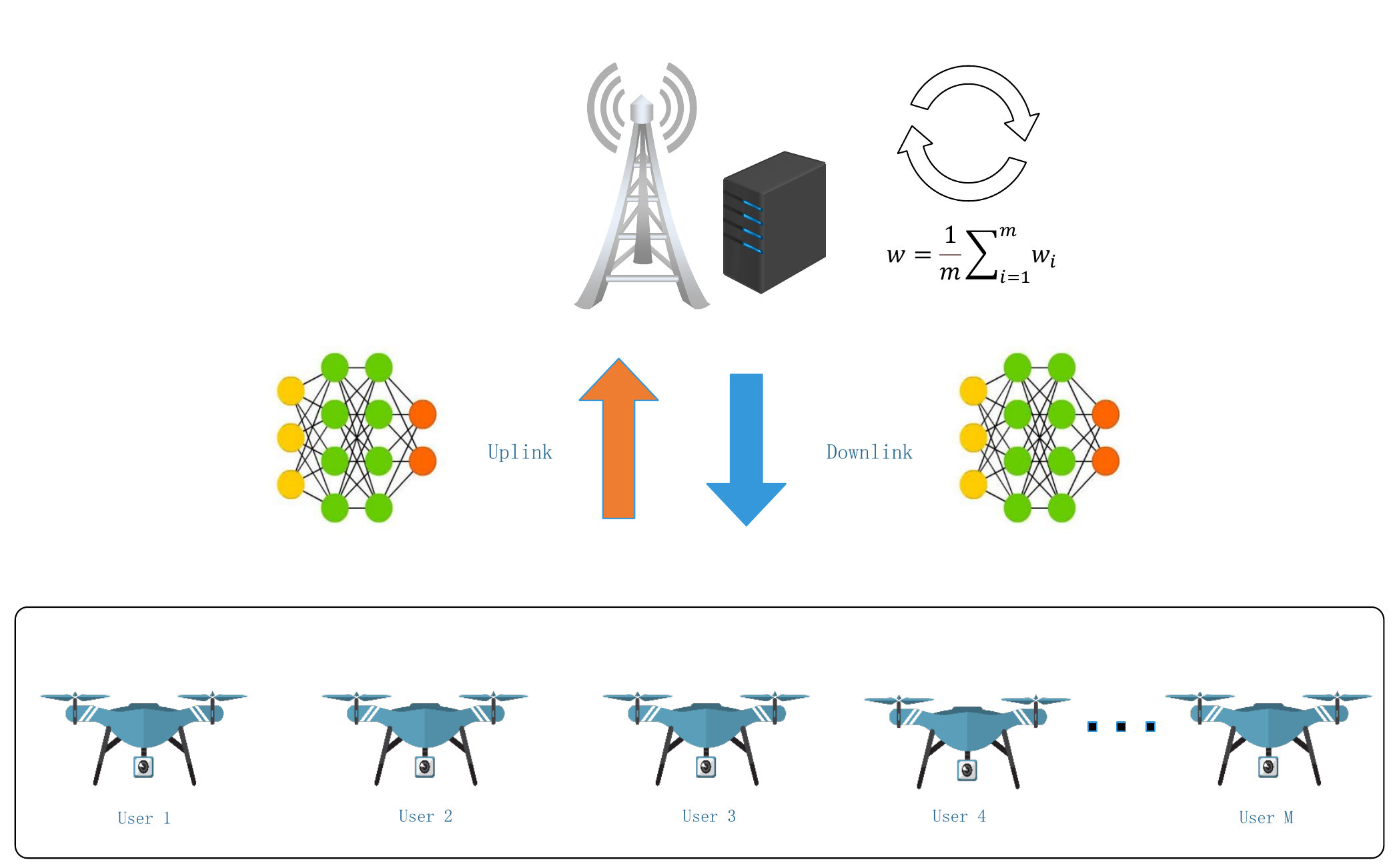}\vspace{-3mm}
    \caption{System model of FEEL in the UAV-enabled IoT networks.}\vspace{-5mm}
    \label{FL}
  \end{figure}
\section{System Model and Problem Formulation}
  
  \subsection{Federated Learning}
  Federated learning is a distributed machine learning method to train a shared model in the context of protecting personal privacy.
  It allows users to train their dataset locally instead of uploading the sensitive data to a server. The server collects the information 
  from different users and generates a global model. This process can be expressed as
  
  \begin{align}\label{FL_opti}
   \arg\min_{w \in \mathbb{R}}  F(\omega)=\frac{1}{M}\sum_{m=1}^M F_m(\omega),
  \end{align}
  where $\omega$ is the model weight, $M$ is the  user number, and $F_m(\cdot)$ is the loss function. 
  As a matter of fact, FedAvg\cite{FL2016} is used to aggregate
  a global model to reduce the communication rounds. Firstly, all the devices train their models locally and then update the weight as

 \begin{align}\label{FL_update}
  \omega_{k+1}^m \leftarrow \omega_{k}^m- \alpha \nabla F_m(\omega),
  \end{align}
where $\alpha$ is a positive learning rate, $\nabla(\cdot)$ denotes the gradient operation and $k$ is  the round index. Each user executes (\ref{FL_update}) several times, and then it uploads the weight $w_{k+1}^m$
to the server. The server aggregates the collected weights and calculates the weighted average weights as
\begin{align}\label{FL_avg}
  \overline{\omega}_{k+1} \leftarrow \sum_{m=1}^M \frac{n_m}{n}\omega^m_{k+1},
\end{align}
where $\frac{n_m}{n}$ is the proportion of samples in total. It is called as model average\cite{average1,average2_aaai}.
Lastly, the server broadcasts the new global model to each user.

\subsection{System Model}
  Fig. \ref{FL} shows the system model of the considered FEEL framework, where there are $M$ distributed users and a centralized server.
  In practical, the $M$ users can be various UAV devices in the edge computing-based IoT network, and the parameters server can be the base station or one of the UAVs. 
  We use $\mathcal{M}=\left\{1,2,\cdots,M\right\}$ to denote the user set.
  The number of data samples which needs to be trained locally is $d_m$.
  We assume that a round of FL training can be completed in a time slot. For each device $m \in \mathcal{M}$ in the $k$-th round, 
  the base CPU-frequency is $f_m^k$ and it requires $c_m$ CPU-cycles to process a sample of the local dataset. 
  To deal with the problem of limited communication resources, each device will train $e$ times locally before uploading the weights\cite{FL2016}.
  The local training time of device $m$ at the $k$-th round can be written as
  
  \begin{align}\label{training_time}
    t^k_{m,local}=\frac{e c_m D_m}{\eta_m^k f_m^k},
  \end{align}
  where $\eta_m^k$ is the coefficient of frequency adjustment which determines the practical operating CPU-frequency
  and $K$ is the maximum round of FL.
  The weights will be uploaded to the centralized server through the wireless link after each device completes the local training.
  The  data rate of the wireless link from device $m$ to the centralized server is:

  \begin{align}\label{Shannon}
  r_{m}^k = B_{m}^k \log_2 \left(1 + \frac{P_{m} |h_m^k|^2}{\sigma_{m}^2}\right),
  \end{align}
  where $h_m^k \sim \mathcal{CN}(0,\beta)$  is the channel parameter of the link from device $m$ to the centralized server, 
  $\sigma^2$  is the variance of the additive white Gaussian noise (AWGN) at the server and $P_m$ is the transmit power of device $m$. 
  Notation $B_{m}^k$ is the bandwidth of device $m$ and it can be allocated by the base station. At each round, $B_m^k$ should
  meet the following requirement
  \begin{align}\label{bandwidth_sum}
    \sum_{m=1}^M B_m^k=B_{total}.
  \end{align}
  According to (\ref{Shannon}), the transmission latency can be calculated as
  \begin{align}\label{transmission_time}
    t^k_{m,up}=\frac{\epsilon}{r_{m}^k}, 
  \end{align}
  where $\epsilon$ is the size of deep learning model's weights. 
  The total time to provide a local model of device $m$ can be expressed as
  \begin{align}\label{total_latency}
    T^k_{m}=t^k_{m,local}+t^k_{m,up}.
  \end{align}
  Each device  trains its local model and uploads parallelly. The centralized server has to wait for the local model from the slowest device which can be viewed as the bottleneck of the system to aggregate 
  a new global model. So the system latency at the $k$-th round is
  \begin{align}\label{system_latency}
    T_k=\max_{m \in \mathcal{M}} {T^k_{m}}.
  \end{align}
  
  Similarly, we can calculate the energy consumption in two stages. At the first stage, device $m$ performs 
  some algorithms like back propagation (BP) to update the weights at the cost of huge energy consumption. According to [2], the training energy 
  consumption can be written as
  \begin{align}\label{total_latency}
    E^k_{m,local}=\zeta_m c_m D_m (\eta_m^k f_m^k)^2,
  \end{align}
  where $\zeta_m$ is the energy consumption coefficient of CPU chips. Then device $m$ uploads the weights with transmit power $P_m$ 
  and the energy consumption of transmission can be easily expressed as
  \begin{align}\label{uploading_energy}
    E^k_{m,up}=P_mt^k_{m,up}.
  \end{align}
  So the total energy consumption of device $m$ at the $k$-th round is
  \begin{align}\label{total_energt}
    E^k_m=E^k_{m,local}+E^k_{m,up}.
  \end{align}
  Under ideal conditions, devices can continuously participate in FedAvg\cite{FL2016} and contribute their local models 
  until achieving the best performance. However, in the case of limited resources, we have to consider that 
  the battery of devices may run out. The energy model should meet the following requirement
  \begin{align}\label{Battery_constrained}
    \sum_{k=1}^K E^k_m \leq \delta_m,
  \end{align}
  where $\delta_m$ is the total battery power of device $m$.
  
  To describe the total system cost, we use a linear combination of latency and energy consumption as following \cite{zhan2020experience,lichao}
  \begin{align}\label{linear combination}
    \Phi=\sum_{k=1}^{K}\left[\lambda T_k +(1-\lambda)\sum_{m}^M E^k_m\right],
  \end{align}
  where $\lambda \in \left[0,1\right]$ is a factor to describe the importance of latency and energy consumption in the system cost.
  We can adjust $\lambda$ to trade-off the latency and energy consumption. 
  Specifically, the linear combination approaches to the  energy consumption when $\lambda$ goes to zero, while it degenerates into the latency if $\lambda$ is near one.
  In fact, the linear combination is a method to solve  multi-objective programming 
  problems by giving a proper weight coefficient according to the importance of each objective and changing to
  single-objective programming problems.

\subsection{Problem Formulation}
For the practical battery-constrained devices, we expect that they can participate in the rounds of FEEL training as much as possible  and meanwhile try to avoid running out of power.
Moreover, the system cost measured by the latency and energy consumption should be minimized, which is particularly important for the IoT devices. 
By taking into account these factors, we can optimize the training rounds and system cost by adjusting the  CPU-frequency and bandwidth. The system optimization problem can be given by
\begin{align}\label{Problem Formulation}
  \max_{\{ {\eta_m^k},B_m^k\}} K-\ \overline{\Phi}   & \nonumber \\
  \quad \quad \mathrm{s.t.} \quad
  C_1:  &  \eta_m^k \in [0,1], \\
  C_2:  & \sum_{m=1}^M B_m^k=B_{total},\nonumber \\
  C_3:  &  \sum_{k=1}^K E^k_m \leq \delta_m,\nonumber 
\end{align}
where $\overline{\Phi}=\frac{\Phi}{K}$ is the average cost of per round. It is generally very hard to solve the above optimization problem by using 
conventional optimization methods such as convex optimization. Especially, in a time-varying environment, the base CPU-frequency
and channel condition are different in each round, causing heterogeneity in the training process. Due to these reasons, 
a learning based algorithm should be developed to adapt to different states in order to find a proper solution. The notations of this section we have used are summarized in Table. \ref{bs2}.

\begin{table}[t]
  \caption{Symbol notations}
  \vspace{20pt}
  \centering
  \begin{tabular}{p{2cm}p{6cm}}
      \hline
      Notation & Definition\\
      \hline
$\alpha$  &  learning rate of FL\\
$\omega^m_{k}$  &  model weight of the $m$-th user in the $k$-th round\\
$\epsilon$ &  size  of  deep  learning  model weight\\
$\zeta_m$ & energy consumption coefficient of the $m$-th user\\
$\Phi$ & system cost\\
$r^k_m$ &transmission rate between the $m$-th user and the server \\
$e$ & local training times\\
 $c_m $ &  cpu-cycles to process a sample of the $m$-th user \\
 $d_m $ &  number to data samples of the $m$-th user \\
 $\delta_m$ & total battery power of device $m$\\
$\eta_m^k$ &  coefficient of frequency adjustment of the $m$-th user in the $k$-th round \\
$f_m^k$   &    base CPU-frequency of the $n$-th user in the $k$-th round \\
$t^k_{m,local}$& local latency of the $m$-th user in the $k$-th round\\
$t^k_{m,up}$& transmission latency of the $m$-th user in the $k$-th round\\
$B_m^k$   &    wireless bandwidth of the $n$-th user in the $k$-th round\\
$T^k$& system latency in the $k$-th round\\
$E^k_{m,local}$  & Local energy consumption of the $m$-th user in the $k$-th round\\
$E^k_{m} $ & system energy consumption of the $n$-th user in the $k$-th round\\
$F(\cdot)$&  loss function of FL\\

       \hline
  \end{tabular}
  \label{bs2}
\end{table}

\section{DDPG-Based Allocation Strategy}
In this section, we will try to solve the system optimization problem in (\ref{Problem Formulation}), by using the DDPG-based allocation strategy. 
Specifically, we will first describe the Markov decision process (MDP), and then introduce how to implement the DDPG-based resource allocation strategy.
\subsection{Markov Decision Process}
MDP is used to model decision-making in time-varying environments and it mainly consists  of a 4-tuple $\{S,A,P_a,R_a\}$. Specifically, $S=\{k,\mathbf{O^k},
\mathbf{F^k},\mathbf{H^k}\}$ is the state space, where $\mathbf{O^k}$ is a vector of the remaining battery power, $\mathbf{F^k}$ is a vector of the current base CPU-frequency of all the devices, and $\mathbf{H^k}$ is the 
channel parameter of the $k$-th round which can be obtained by using some channel estimation methods. We use $A=\{\mathbf{N^k},\mathbf{B^k}\}$ to denote 
the action space, where $\mathbf{N^k}=\{\eta_1^k,\eta_2^k, \cdots \eta_M^k \}$ is the coefficient of CPU-frequency adjustment 
and $\mathbf{B^k}=\{B^k_1,B^k_2 \cdots B^k_M\}$ is the allocated vector of bandwidth. 

At the $k$-th round of the FEEL training, the current state is $s_k \in S$ and the agent will perform an action $a_t \in A$ in the environment.
From the feedback of environment, $s_k$ will transit to $s_{k+1}$ with a conditional probability $P$. Meanwhile, the agent will 
achieve an instant reward from the environment, which can be expressed by
\begin{align}\label{linear combination}
  R^k=k-\phi,
\end{align}
where $k$ is positive feedback and $\phi$ is the instant cost of a round in FEEL training, which is denoted by the linear combination 
of energy consumption and latency. The agent expects to achieve a long-term average reward from the environment by maximizing $K$ and minimizing 
$\overline{\Phi}$.

However, it is difficult to estimate the conditional probability $P$ in many application scenarios. Hence, we turn to use the DDPG algorithm to solve this problem.
\begin{figure}[tbp]
  \centering
  \includegraphics[width=3.5in]{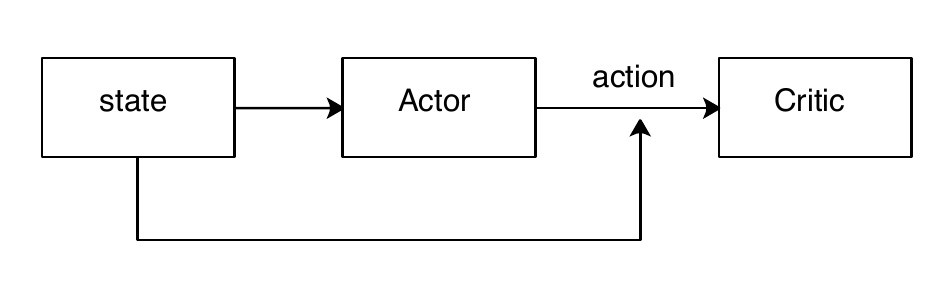}\vspace{-3mm}
  \caption{Actor-critic framework.}\vspace{-5mm}
  \label{actor-critc}
\end{figure}
\subsection{Deep Deterministic Policy Gradient}

In the considered FEEL scenario, the CPU-frequency and bandwidth allocation vary over a continuous range with infinite possibilities. 
Accordingly, although deep Q-learning network \cite{DQN1,DQN2} can deal with the problem of discrete and low-dimensional action space, it is unable to work
efficiently in the face of continuous action space since its performance severely relies on finding the maximum of value function approximated by the neural network in each iteration. 
Due to this reason, we turn to use the DDPG strategy to solve the optimization problem of CPU-frequency and bandwidth allocation. Deterministic Policy Gradient\cite{DPG} was proposed to output a deterministic action, 
instead of the value of all the actions under the actor-critic framework. Fig. \ref{actor-critc} shows the actor-critic framework in the DDPG, 
where a deep neural network called actor network is used to produce a deterministic
action while a critic network is employed to approximate the value-function
which evaluates the action. 
To formulate the DDPG strategy, we use $\mu(s|\theta^{\mu})$ and $Q(s,a|\theta^{Q})$ 
to denote the actor and  critic network, respectively. 
For a deterministic action $a_t$ in the state $s_t$, the Bellman equation for the action-state value function can be written as
\begin{align}\label{Q-value}
  Q(s,a)=\mathbb{E}_{s_{t+1}\sim S} \left[r(s_t,a_t)+\gamma Q(s_{t+1},a_{t+1})\right],
\end{align}
where $s_{t+1}$ is the next state when the agent executes the action $a_t$ in the state $s_t$ and $a_{t+1}$ is the next action given by the actor network.

\begin{figure}[btp]
  \centering
  \includegraphics[width=3.5in]{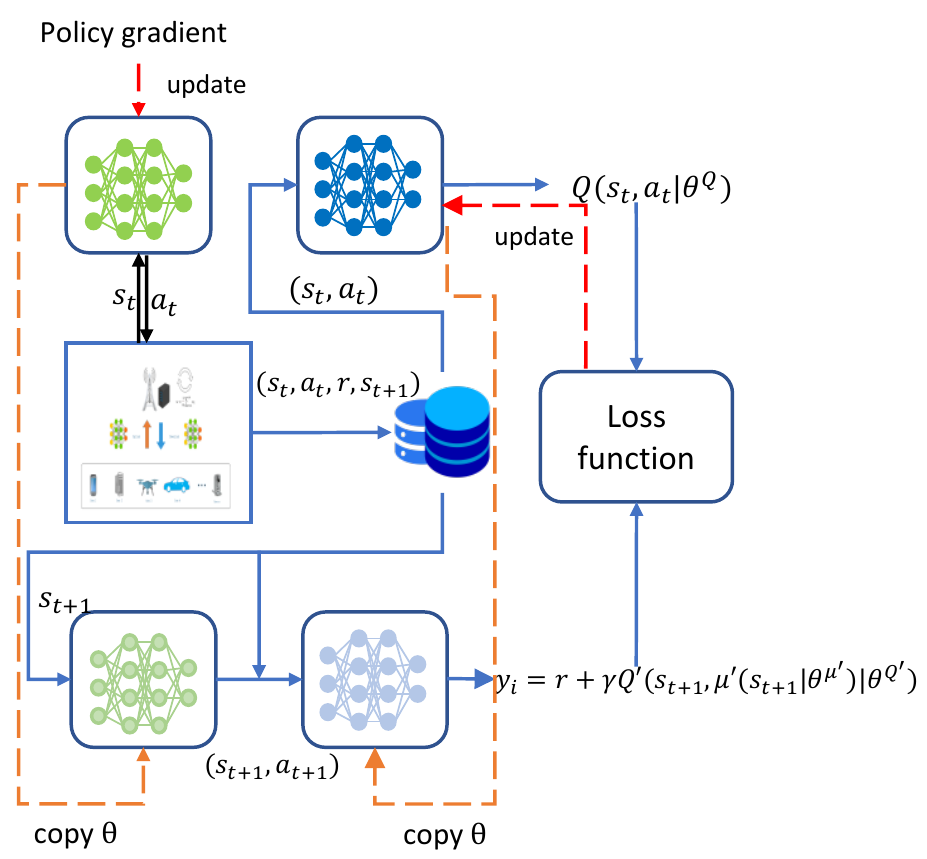}\vspace{-3mm}
  \caption{Implementation structure of DDPG.}\vspace{-5mm}
  \label{DDPG}
\end{figure}

Motivated by the idea of double network in the DQN, target networks and main networks are used in DDPG. Fig. \ref{DDPG} shows the implementation structure of DDPG, where there are four neural
networks working together,
\begin{itemize}
  \item \textbf{Main actor network $\mu$}: We will get the deterministic action from $a_t=\mu(s|\theta^{\mu})$ and $\mu$ is used to update weights 
  of the target actor network.
  \item  \textbf{Target actor network $\mu'$}: The target actor network predicts the next action by $a_{t+1}=\mu'(s_{t+1}|\theta^{\mu'})$.
  \item \textbf{Main critic network $Q$}: According to the current state $s_t$ and action given by the main actor network $\mu$, the main critic network $Q$
  is responsible for giving the $Q$ value by $Q(s,a|\theta^{Q})$.
  \item \textbf{Target critic network $Q'$}: In the state $s_{t+1}$, the $Q$-value will 
  be evaluated by $Q'(s_{t+1},\mu'(s_{t+1}|\theta^{\mu'})|\theta^{Q'})$.
\end{itemize}
The approach of target networks reduces the correlation between the current $Q$-value and target $Q$-value, which helps increase the robustness of training.
For the main critic network, the loss function can be expressed as
\begin{align}\label{loss_critic}
  L(\theta^Q)=\mathbb{E}_{\mu'}(Q(s,a|\theta^{Q})-y_i)^2,
\end{align}
where
\begin{align}\label{yi}
  y_i=r(s_t,a_t)+\gamma Q'(s_{t+1},\mu'(s_{t+1}|\theta^{\mu'})|\theta^{Q'}),
\end{align}
in which $\gamma$ is a positive discount factor of reward. By executing the gradient descent method and minimizing the loss function in (\ref{loss_critic}), we can update the main critic network.
Additionally, to optimize action decisions and achieve higher rewards, $Q$-value is considered  as the loss function of the actor network.
The main actor network will be updated by the gradient ascent from
\begin{align}\label{gradient_actor}
  \nabla_{\theta^{\mu}}=\mathbb{E}_{\mu'}\left[\nabla_a Q(s,a|\theta^{Q})|_{s=s_t,a=\mu(s_t)} \nabla_{\theta^{\mu}} \mu(s|\theta^{\mu})|_{s_t}\right].
\end{align}
Different from copying weights from the main networks in a certain interval of time in DQN, the target networks in DDPG adopt a soft update in each step 
as
\begin{align}\label{soft_update}
  \theta ^{Q'} \leftarrow \tau \theta +(1-\tau)\theta^{Q'} \\
  \theta ^{\mu'} \leftarrow \tau \theta +(1-\tau)\theta^{\mu'},
\end{align}
where $\tau$ is an update factor. In addition, an experience reply unit (ERU) is used to collect training samples and the agent will randomly choose batches of samples during the training 
process. This can help break the relevance and non-stationary distribution between training samples and improve the performance.

\subsection{DDPG-based resource allocation strategy}
\begin{algorithm}[tbp]
  \caption{DDPG-based resource allocation strategy}
 
  \label{DDPG_algor}
  {\textbf{Input:}} {current state $S=\{k,\mathbf{O^k},\mathbf{F^k},\mathbf{H^k}\}$}\\
  {\textbf{Output:}} {allocation matrix $A=\{\mathbf{N^k},\mathbf{B^k}\}$}
\begin{algorithmic}[1]

\State Initialize the target network $Q'$ and $\mu'$ with $\theta^{Q'} \leftarrow \theta^Q$,$\theta^{\mu'}\leftarrow \theta^{\mu}$ in the server.
\State Initialize experience replay memory $\mathcal{D}$
  \For{episode = 1,M}
  \State Initialize a random process $\mathcal{N}$ for exploring.
  \State Initialize state $s_1$
   \While{$s \notin S_{end}$}
  \State Devices upload the base frequency to the server and estimate the channel parameters.
  \State  Server chooses an action $a_k=\mu(s_t|\theta^{\mu})+\mathcal{N}_t$
  \State  Execute action the $a_k$ and observe reward $r_k$ and next state $s_{k+1}$
  \State  Store $(s_k,a_k,r_k,s_{k+1})$ in $\mathcal{D}$
  \State  Randomly sample a mini-batch of  transitions $(s_i,a_i,r_i,s_{i+1})$ from $\mathcal{D}$
  \State $  y_i=r(s_i+\gamma Q'(s_{i+1},\mu'(s_{i+1}|\theta^{\mu'})|\theta^{Q'}).$
  \State Update the critic network by minimizing :
  $$L=\frac{1}{N}{\mu'}(Q(s,a|\theta^{Q})-y_i)^2$$
  \State Update the actor network by gradient ascent: 
  $$\nabla_{\theta^{\mu}}=\mathbb{E}_{\mu'}\left[\nabla_a Q(s,a|\theta^{Q})|_{s=s_t,a=\mu(s_t)} \nabla_{\theta^{\mu}} \mu(s|\theta^{\mu})|_{s_t}\right]$$
  \State Update the target networks: 
  $$ \theta ^{Q'} \leftarrow \tau \theta +(1-\tau)\theta^{Q'}$$
  $$\theta ^{\mu'} \leftarrow \tau \theta +(1-\tau)\theta^{\mu'}$$

   \EndWhile
  \EndFor

\end{algorithmic}
\end{algorithm}
In this part, we will introduce the proposed DDPG-based resource allocation strategy. Firstly, we design a mutil-task neural network
which outputs the results of $N_k$ and $B_k$ simultaneously. In order to ensure the outputs  in the correct range, the Sigmoid and Softmax activation functions
are used in the last layer of network. Before each round of FEEL training, clients need to report their available resources such as the base CPU-frequency and 
remaining battery power to the server. At the same time, the server should estimate the channel parameters of the links from it to the users.
These information is entered as the input of the main actor network to get a deterministic action of adjusting
the practical CPU-frequency and bandwidth allocation. The base station  will inform participants the operating CPU-frequency and the available wireless bandwidth
to them. The agent will achieve different rewards in each round according to the latency, energy consumption and the current number of rounds. 
When the experience reply unit has enough samples, the agent will adjust its strategy by the way introduced in the previous
part. In order to guarantee the global convergence, we reset the environment and re-initialize
state $s_1$ to train the agent when the agent reaches the final state. 
After some episodes with convergence, the DDPG strategy can find a proper result of $\eta_m^k$ and $B_m^k$. In this way, the optimization problem of (\ref{Problem Formulation}) is solved. The whole procedure of the DDPG strategy is summarized in Algorithm \ref{DDPG_algor}.

\section{Simulations and Discussions}
In this section, we will demonstrate the performance of the proposed DDPG-based resource allocation strategy by simulations. 
To simulate the practical scenarios of FEEL, we suppose that there are 20 users with limited battery power subject to $\mathcal{U}(2\times 10^4,
3\times 10^4)$ and their initial base CPU-frequencies are subject to $\mathcal{U}(1\times 10^7,5\times 10^7)$. 
The users will train locally 5 times per round and participate in 1000 rounds of the FEEL training at most. 
In addition, the number of CPU-cycles to process a sample is set subject to $\mathcal{U}(7 \times 10^4 ,2\times 10^5)$, the sizes of users' dataset are subject to $\mathcal{U}(400\\,600)$, and the energy consumption coefficients of CPU chips are set subject to $\mathcal{U} (1\times 10^{-22},2\times 10^{-22})$.
Moreover, the transmit power of all the devices 
is set to $5\times 10^{-5}$, the total wireless bandwidth is $5$MHz, and the model size is set to $10$MB. We assume that the condition will be static in a round,
while it is time-varying in different rounds of training. To simulate the time-varying environments, we set the initial base CPU-frequency varying from $0.8$ to $1.2$ times. 
Similarly, the average channel gain of the wireless links is set to unity, and the variance of the AWGN is set to $10^{-9}$. 

As to the DDPG network, we implement it by the well-known Pytorch library \cite{pytorch}. The actor networks consist of 2 hidden layers with 64 and 256 nodes, respectively. In the critic networks,
there are 2 hidden layers with 30 nodes per layer. To enhance the fitting ability of the networks, the Rectified Linear Unit (ReLU)\cite{relu} is used as
the activation function. The Adam\cite{adam} method is used to optimize the loss function of networks and the learning rates are $10^{-6}$ and $10^{-2}$ in the main actor and main critic networks, respectively.
We set the capacity of ERU to $10^4$ and the batch size is $128$.
Besides, the value of $\gamma$ is $0.999$, and $\tau$ is set to $10^{-3}$. The DDPG agent will train $50$ times at the end of FEEL in each episode.
The total episode of DRL is $800$, and we repeat the experiments to reduce the accidental error.

\begin{figure}[tbp]
  \centering
  \includegraphics[width=3.5in]{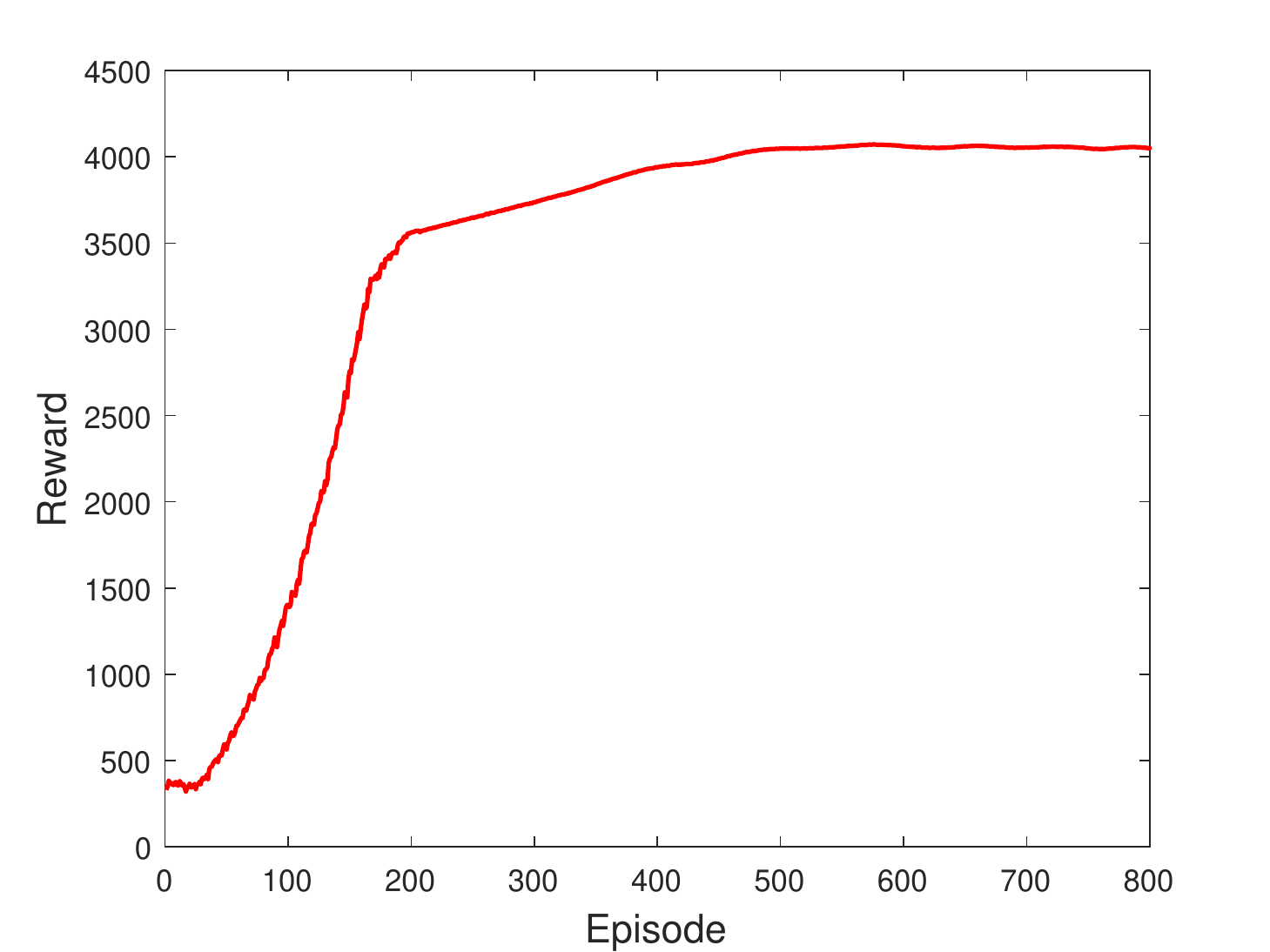}\vspace{-3mm}
  \caption{Convergence of the proposed DDPG approach with $\lambda = 0.5$.}\vspace{-5mm}
  \label{reward}
\end{figure}
\begin{figure}[tbp]
  \centering
  \includegraphics[width=3.5in]{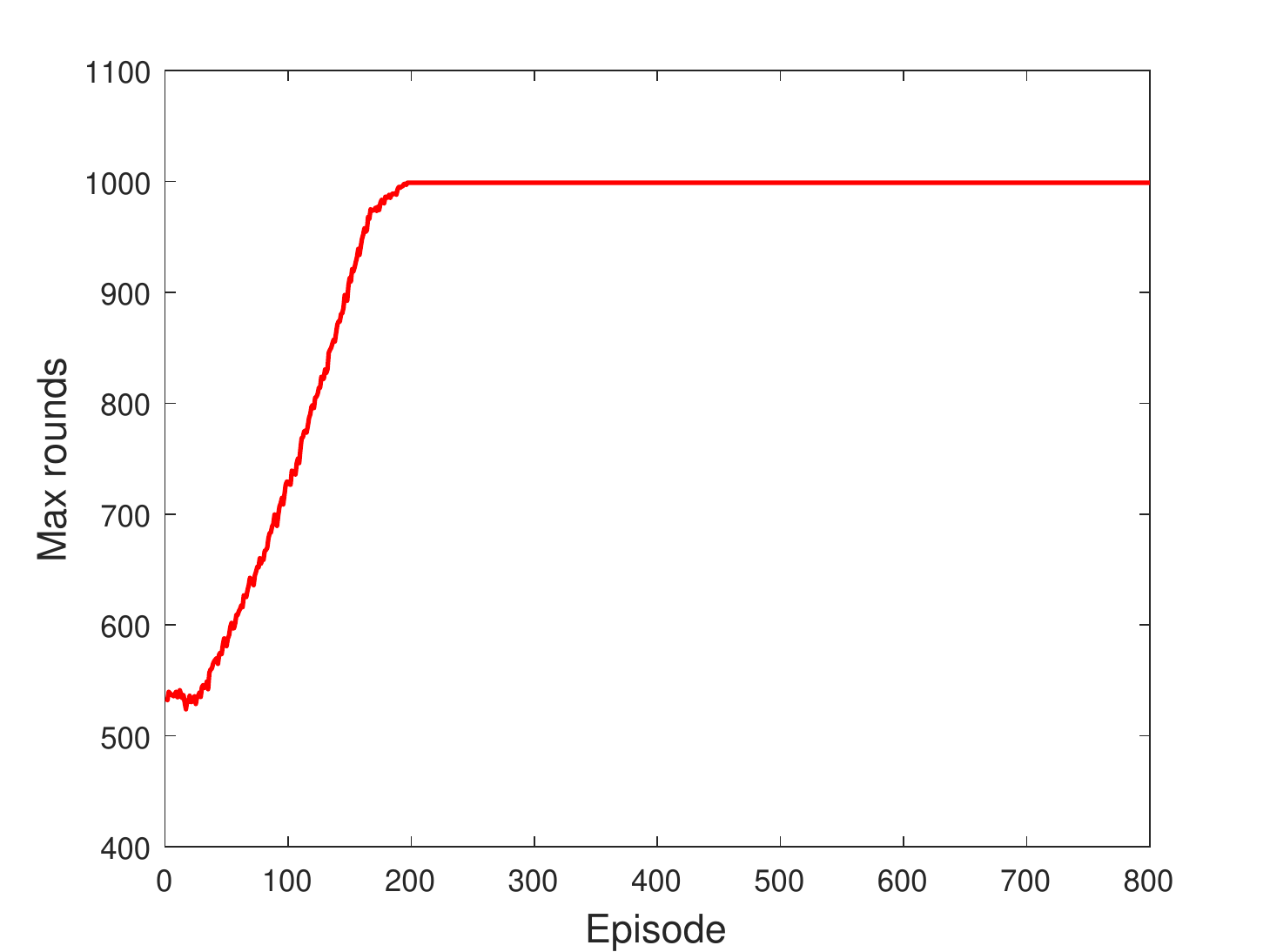}\vspace{-3mm}
  \caption{Convergence of the maximum round.} \vspace{-5mm}
  \label{round_ep}
\end{figure}

Fig. \ref{reward} shows the training process of the proposed DDPG-basd resource allocation strategy, where $\lambda=0.5$ and there are 800 episodes at all.
We focus on the total reward in each episode. From this figure, we can find that the curve of the total reward grows very fast in the first 
200 episodes and then it continues to increase with a slow growth tendency. After about 500 episodes, 
the proposed DDPG approach can achieve an asymptotic reward value of about 4000. These results can help verify the proposed
DDPG-based strategy.

Fig. \ref{round_ep} depicts the maximum round of the proposed strategy versus episodes, where $\lambda=0.5$ and there are 800 episodes at all. From this figure, we can observe that 
at the beginning of episode, the battery-constrained devices are easily to run out of power and can only participate in about 500 rounds.
When the episode increases, the devices can participate in more rounds of training. This is because that with the learning of
DDPG agent, the proposed resource allocation strategy continues to improve and all the devices can prolong the life of their batteries.
When the curve is converged, all the devices can complete 1000 rounds of training. This can not only ensure the performance of training, but also avoid the danger caused by shutdown.

\begin{figure}[tbp]
  \centering
  \includegraphics[width=3.5in]{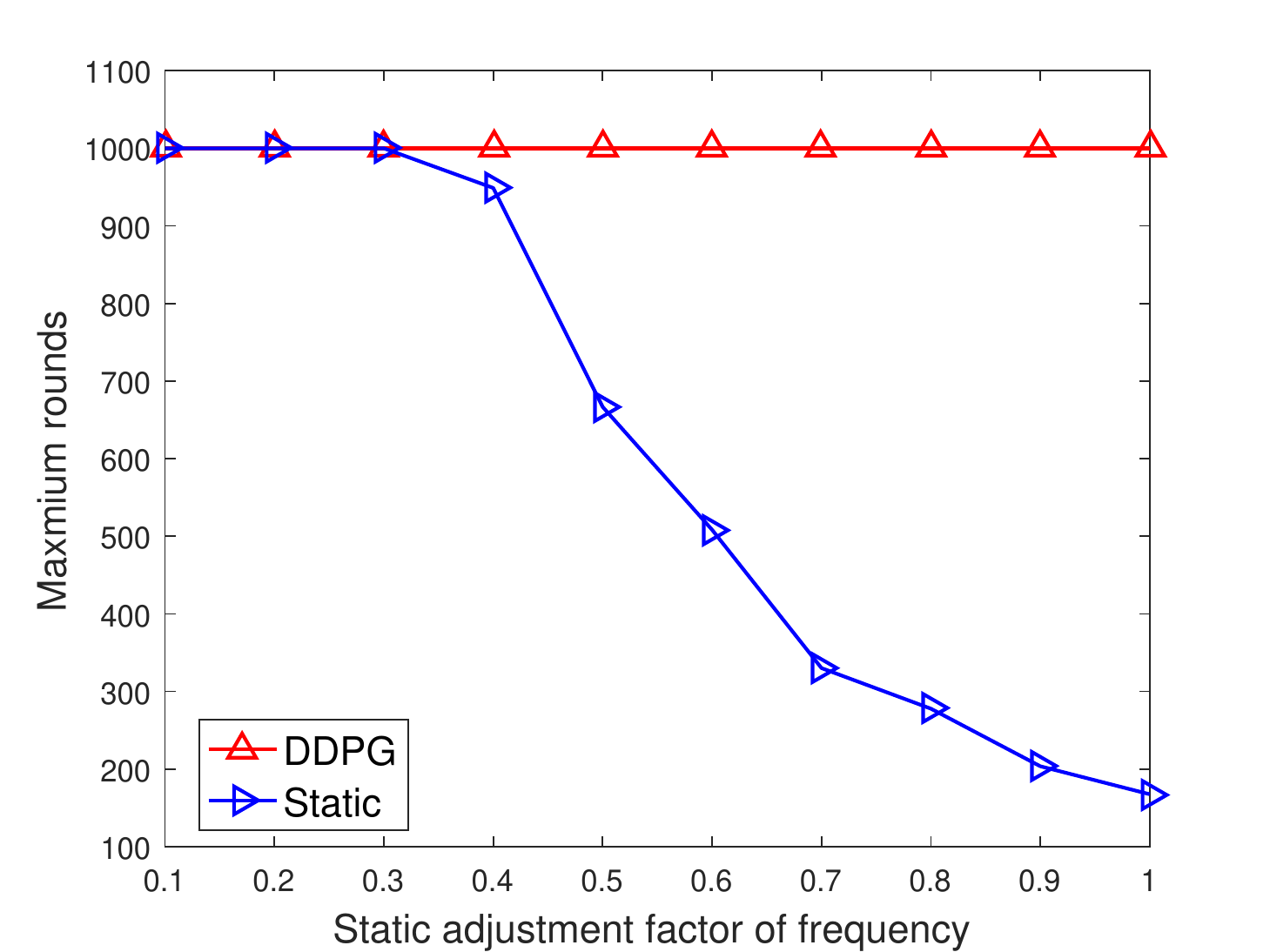}\vspace{-3mm}
  \caption{ Maximum rounds versus the static adjustment factor of frequency.}\vspace{-5mm}
  \label{cmp_round}
\end{figure}

Fig. \ref{cmp_round} shows the maximum round of two strategies versus the static adjustment factor of frequency, where $\lambda=0.5$ and 
the adjustment factor varies from 0.1 to 1. 
For comparison, we provide the performance of the static strategy, 
where the coefficient of frequency adjustment is static at each FL training round and bandwidth allocation is even with 
$B_m^k=B_{total} /M$ for each device. We use the static strategy to compare since in the practical time-varying  scenario, 
it is however difficult to determine how to set the proper operating CPU-frequency of devices, and the devices should run 
at a proportion of base CPU-frequency in order to reduce the energy consumption.
From this figure, we can see that the proposed strategy remains unchanged with the adjustment factor, 
and it can enable all the devices to complete 1000 rounds of training.
In contrast, the static strategy can only obtain about 200 rounds of training when the adjustment factor is around 1, since 
all the devices 
operate at the base CPU-frequency with a high energy consumption.
When the adjustment factor decreases, the static strategy can enable all devices to reduce the energy consumption and 
 take part in more rounds of training.
In particular, the static approach can achieve  1000 rounds of training when the adjusting factor is smaller than 0.3.
The comparison results in this figure can further verify the effectiveness of the proposed DDPG strategy.


\begin{figure}[tbp]
\centering
  \includegraphics[width=3.5in]{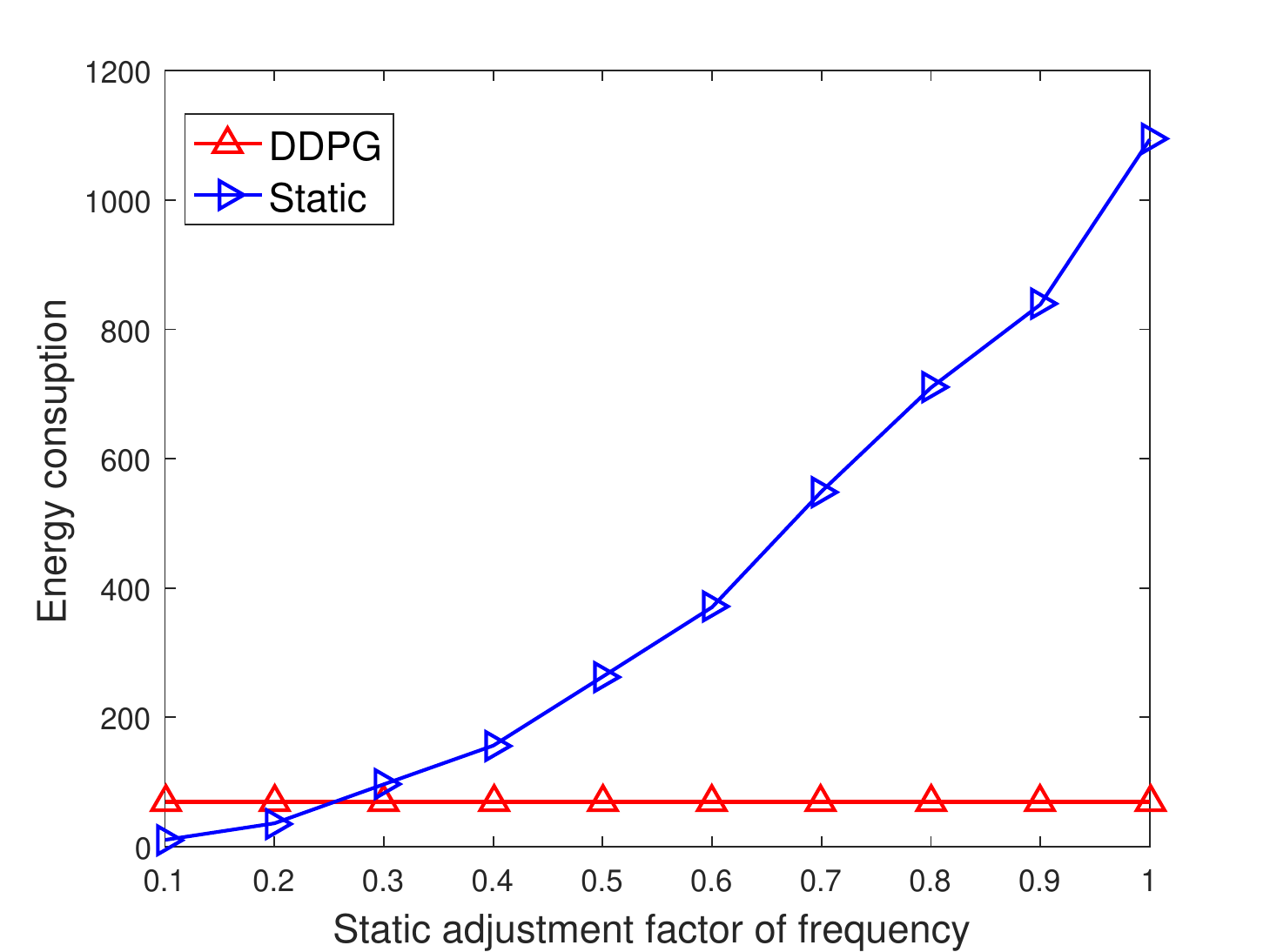}\vspace{-3mm}
  \caption{ Energy consumption versus the static adjustment factor of frequency.}\vspace{-5mm}
  \label{cmp_energy}
\end{figure}

Fig. \ref{cmp_energy} shows the energy consumption versus the static adjustment factor of frequency, where $\lambda=0.5$.
The energy consumption of the proposed strategy is about $69$ and it is a little higher than it in the static approach when
the adjustment factor is smaller than $0.2$. With the increase of the static adjustment factor, the energy consumption 
of the static approach grows explosively. This causes the battery power of devices being easily exhausted and influences the continuous training of FL.

\begin{figure}[tbp]
  \centering
  \includegraphics[width=3.5in]{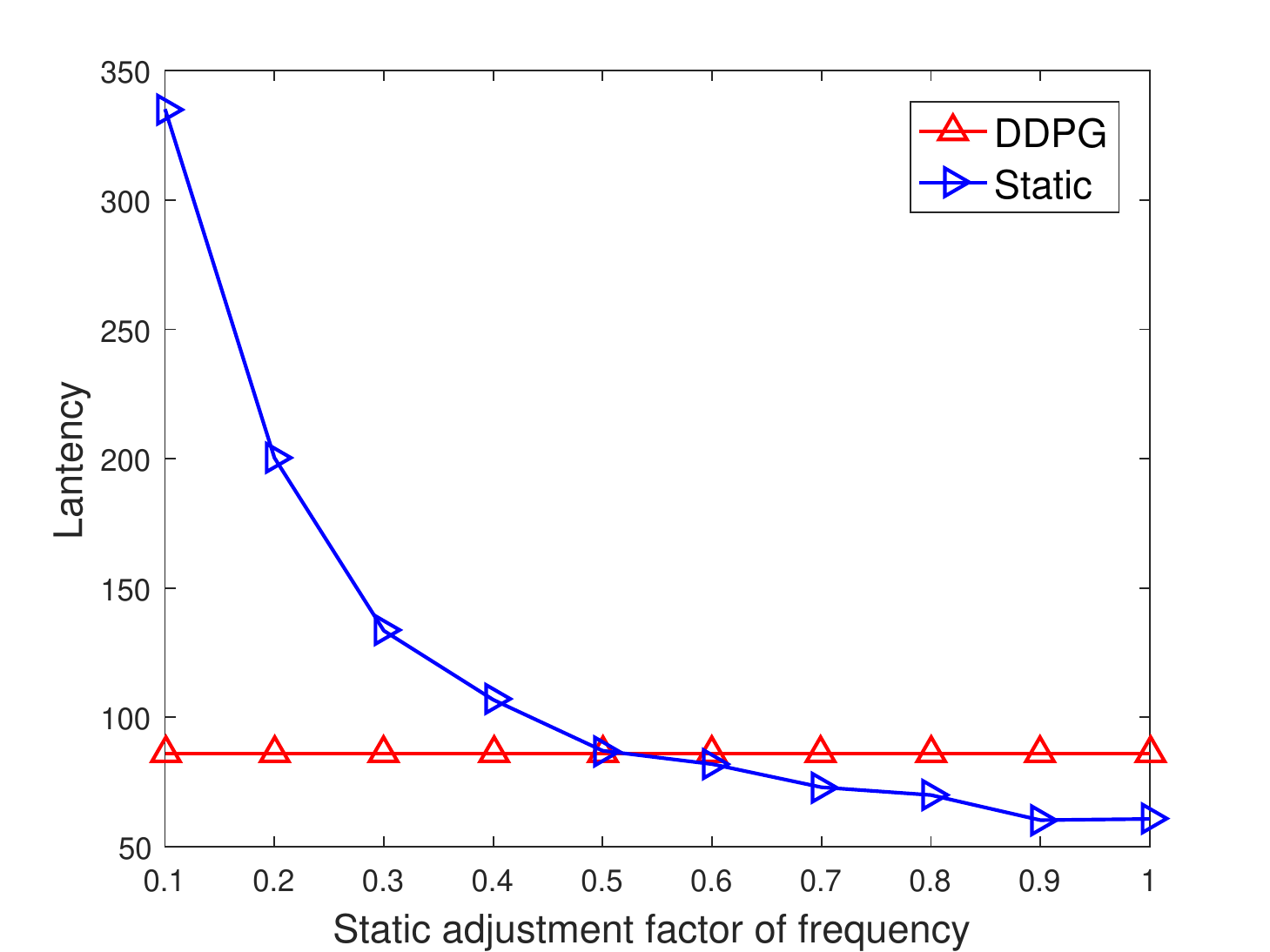}\vspace{-3mm}
  \caption{ Latency versus the static adjustment factor of frequency.}\vspace{-5mm}
  \label{cmp_latency}
\end{figure}

Fig. \ref{cmp_latency} describes  the latency versus the static adjustment factor of frequency, where $\lambda=0.5$.
From this figure, we can observe that the latency is more than 200 when the adjustment factor is smaller than $0.2$ while the latency
of the proposed strategy is only about 96. Hence, it is unacceptable to save energy by decreasing the adjustment factor smaller than $0.2$.
With the increase of the adjustment factor, the latency is lowing down. Moreover, the latency of the proposed strategy is almost identical to that in the static 
strategy when the adjustment factor is $0.5$. By combining the results in Figs. \ref{cmp_energy}- \ref{cmp_latency}, we can conclude that the proposed strategy can make a good trade-off between the energy consumption and latency efficiently. 
This can help verify the effectiveness of the proposed DDPG strategy furthermore.

\begin{figure}[tbp]
  \centering
  \includegraphics[width=3.5in]{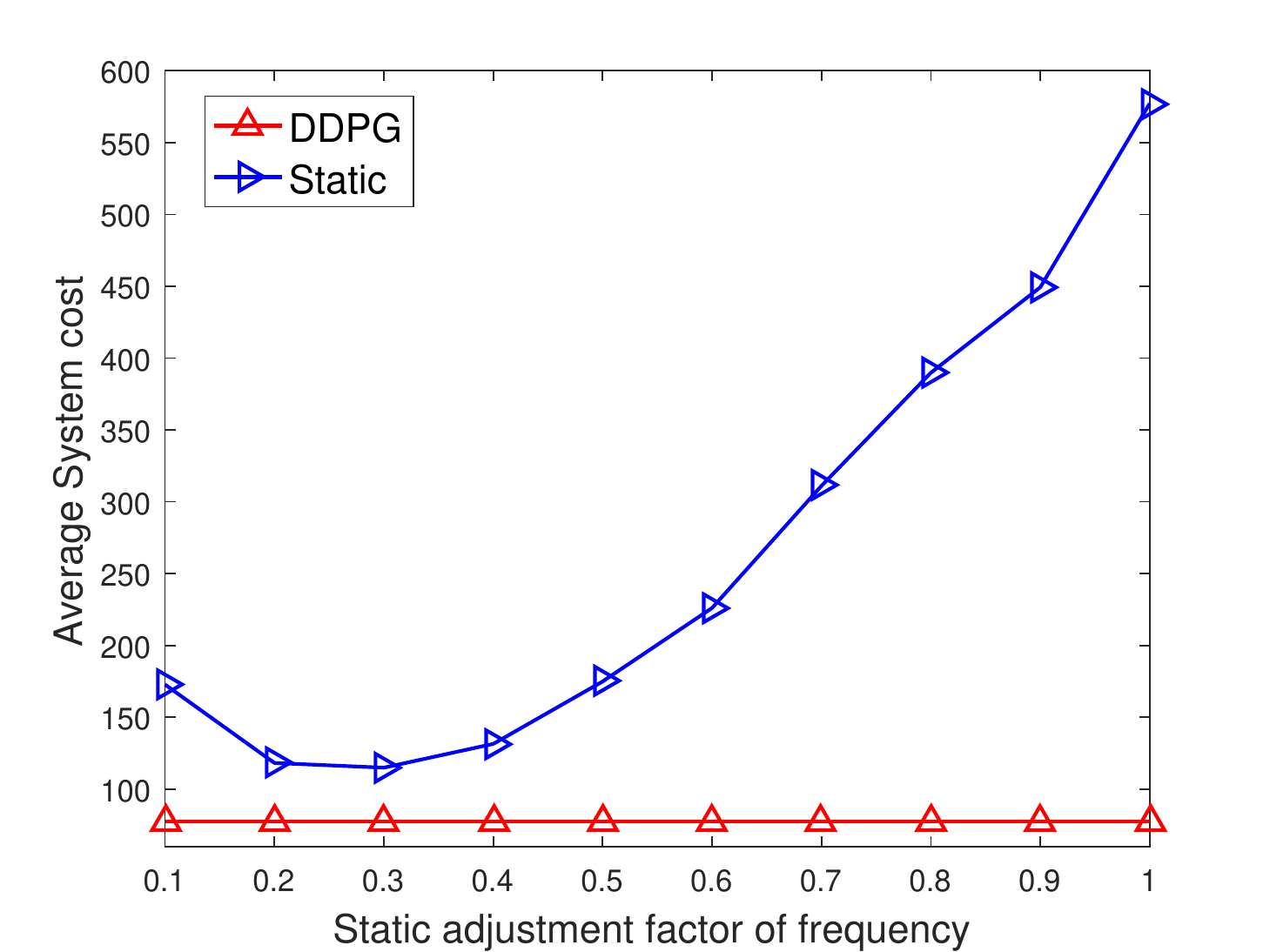}\vspace{-3mm}
  \caption{Average system cost versus the static adjustment factor of frequency.}\vspace{-5mm}
  \label{cmp_LR}
\end{figure}

Fig. \ref{cmp_LR} shows  the system cost versus the static adjustment factor of frequency. To describe clearly, we 
use the linear combination to express the system cost, where $\lambda=0.5$.  
We can find from this figure that the system cost of the proposed DDPG strategy remains unchanged 
with the adjustment factor, which is similar to the phenomena in Figs. \ref{cmp_round}-\ref{cmp_latency}. 
In contrast, the static approach is affected by the adjustment factor significantly. 
Specifically,  the system cost of the static approach becomes smaller when the adjustment factor increases in a low region of adjustment factor. 
This is due to the energy saving, which however leads to a huge latency.
On the contrary,  the system cost of the static approach becomes larger when the adjustment factor increases in a high region of adjustment factor. 
This is because that the pursuit of reducing latency leads to a great energy consumption.
In particular, the static approach achieves the minimum system cost when the adjustment factor is 0.3, which is still higher than that of the proposed DDPG strategy.

\begin{figure}[tbp]
  \centering
  \includegraphics[width=3.5in]{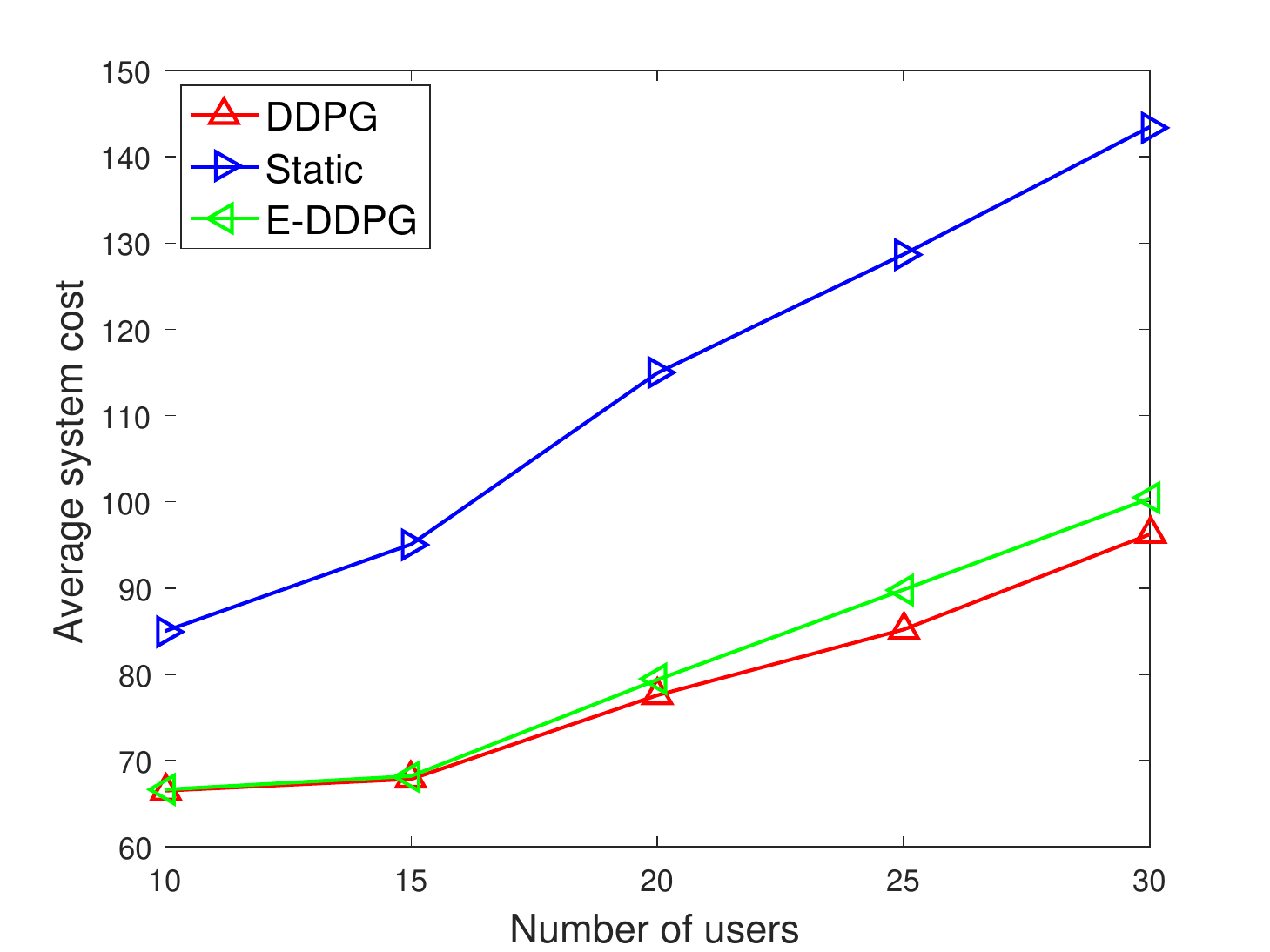}\vspace{-3mm}
  \caption{Average system cost versus number of users.}\vspace{-5mm}
  \label{cmp_users}
\end{figure}

Fig. \ref{cmp_users} depicts the system cost of several strategies versus the number of users, where $\lambda=0.5$. 
For comparison, we plot the performance of the proposed DDPG strategy with even bandwidth allocation, denoted by E-DDPG. 
From this figure, we can find that the proposed DDPG outperforms E-DDPG, since the former can exploit the wireless bandwidth resources in the learning process, 
which can help improve the system performance. Moreover, E-DDPG is much better than the static approach, since the later cannot utilize the system communication and computational resources efficiently. 
In further, the system performance of all the three strategies increases with a larger number of users, as more uses impose a heavier burden on the training process. 

\begin{figure}[tbp]
  \centering
  \includegraphics[width=3.5in]{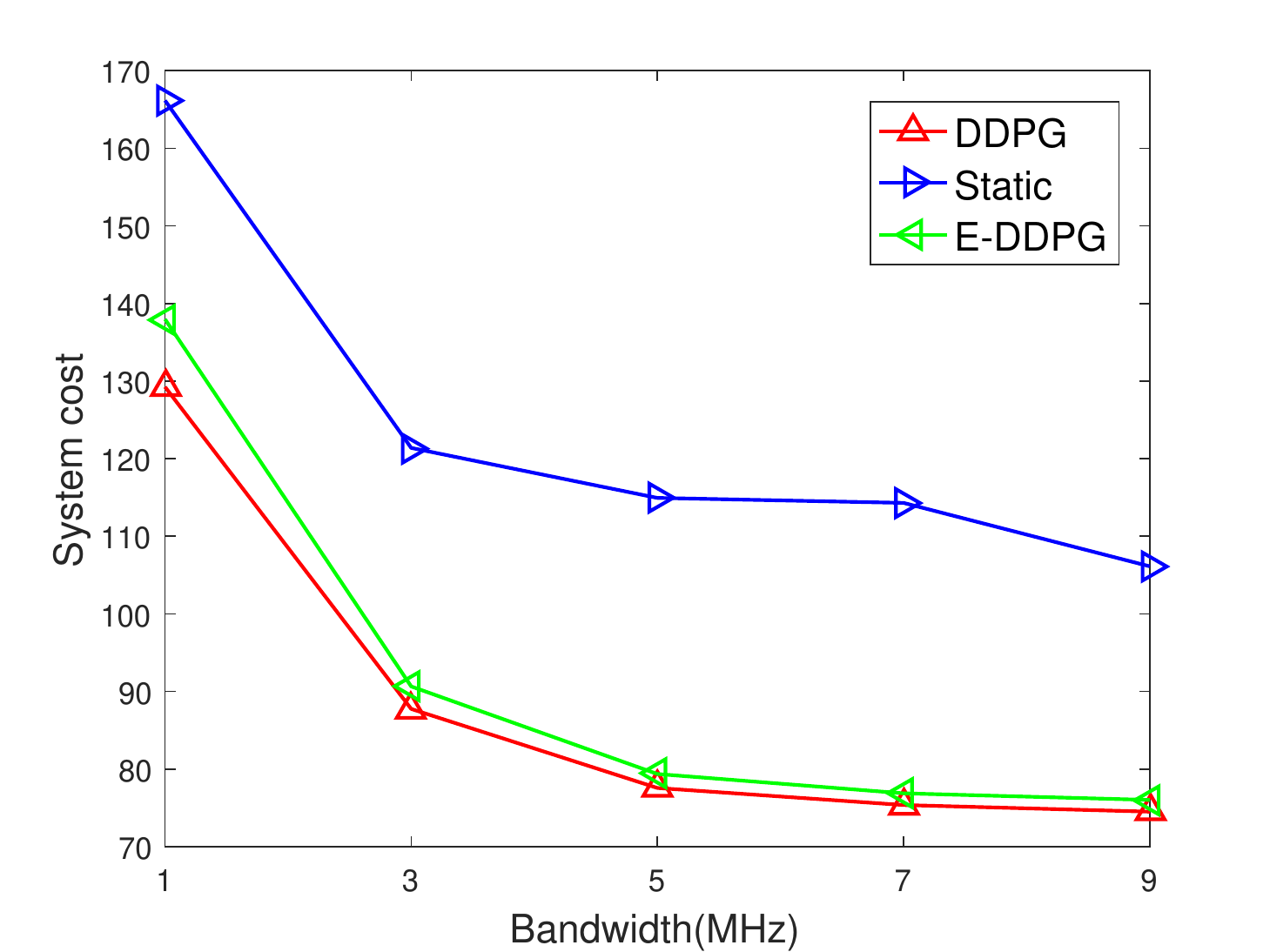}\vspace{-3mm}
  \caption{Average system versus different bandwidth.}\vspace{-5mm}
  \label{cmp_bandwidth}
\end{figure}

In Fig. \ref{cmp_bandwidth}, we plot the performance of several resource allocation strategies versus the wireless bandwidth,
where $\lambda=0.5$ and the wireless bandwidth varies from 1MHz to 9MHz. From this figure, we can see that
the performances of the three strategies become better when the bandwidth increases, since a larger
bandwidth can help reduce the transmission latency as well as the transmission energy consumption.
Moreover, the proposed strategy outperforms the static approach and E-DDPG for various values of wireless bandwidth, 
since it can exploit the system communication and computational resources efficiently. In particular, 
when the total bandwidth is 1MHz, the cost of the proposed DDPG strategy, E-DDPG and the static approach 
is about 130, 140, and 170, respectively. In further, when the bandwidth increases, the transmission latency becomes negligible in the system cost for the proposed DDPG
and E-DDPG strategies, which makes the performance gap between these two strategies decrease. The results in this figure demonstrate the 
merits of the proposed DDPG strategy furthermore.

\section{Conclusions} This paper studied how to optimize the FEEL in UAV-enabled IoT networks where UAVs had limited batteries, from a deep reinforcement learning approach. Specifically, we provided an optimization framework where the devices could adjust their operating CPU-frequency to prolong the battery life and avoid withdrawing from federated learning training untimely, through jointly allocating the computational resource and wireless bandwidth in time-varying environments. To solve this optimization problem, we employed the DDPG based strategy, where a linear combination of latency and energy consumption was used to evaluate the system cost. Simulation results were demonstrated to show that the proposed strategy could efficiently avoid devices withdrawing from the FL training untimely and meanwhile reduce the average system cost.




\bibliographystyle{IEEEtran}

\bibliography{reference}

\end{document}